\documentclass[useAMS,usenatbib]{mn2e}

\usepackage{graphicx}
\usepackage{color}
\usepackage{amssymb}

\usepackage{subfig}
\newcommand{\transpose}[1]{\ensuremath{#1^{\scriptscriptstyle T}}}
\title[Matter Perturbations in Scaling Cosmology]
  {Matter Perturbations in Scaling Cosmology}
\author[A. Romero Fu\~no, W.S. Hip\'{o}lito-Ricaldi and W.~Zimdahl]
  {A. Romero Fu\~no,$^1$\thanks{E-mail: alonsoromero.ufes@gmail.com}
   W.S. Hip\'{o}lito-Ricaldi,$^2$ \thanks{E-mail: wiliam.ricaldi@ufes.br}
   and W.~Zimdahl,$^1$ \thanks{E-mail: winfried.zimdahl@pq.cnpq.br}\\
  $^1$Universidade Federal do Esp\'{\i}rito Santo,
Departamento de F\'{\i}sica, Av. Fernando Ferrari, 514, Campus de Goiabeiras, \\
CEP 29075-910, Vit\'oria, Esp\'{\i}rito Santo, Brazil\\
  $^2$Universidade Federal do Esp\'{\i}rito Santo, Departamento de Ci\^encias  Naturais,
Rodovia BR 101 Norte, km. 60,  Campus\\
de S\~ao Mateus, CEP 29932-540, S\~ao Mateus, Esp\'{\i}rito Santo, Brazil}

\begin{document}

\date{xxxx. xxxx; xxx}

\pagerange{\pageref{firstpage}--\pageref{lastpage}} \pubyear{xxx}

\maketitle

\label{firstpage}

\begin{abstract}
A suitable nonlinear interaction between dark matter with an energy density $\rho_{M}$  and dark energy with an energy density $\rho_{X}$ is known to give rise to a non-canonical scaling $\rho_{M} \propto \rho_{X}a^{-\xi}$ where $\xi$ is a parameter which generally deviates from $\xi =3$.
Here we present a covariant generalization of this class of models and investigate the corresponding perturbation dynamics.
The resulting matter power spectrum for the special case of a time-varying Lambda model is compared with data from the SDSS DR9 catalogue.
We find a best-fit value of $\xi = 3.25$ which corresponds to a decay of dark matter into the cosmological term.
Our results are compatible with the $\Lambda$CDM model at the 2$\sigma$ confidence level. 
\end{abstract}

\begin{keywords}
cosmology: dark energy -- cosmology: dark matter.
\end{keywords}

\section{Introduction}
The currently preferred cosmological model, the $\Lambda$CDM model, is characterized by a pressureless dark-matter (DM) component with an energy density $\rho_{M}$, which decays with the third power of the cosmic scale factor $a$ and a constant energy density $\rho_{\Lambda}$, attributed to a cosmological constant $\Lambda$.
Alternative models, as far as they remain in the context of Friedmann-Lema\^{\i}tre-Robertson-Walker (FLRW) models, replace $\rho_{\Lambda}$ by a (not necessarily constant) $\rho_{X}$, the energy density of dark energy (DE), equipped with an equation-of-state (EoS) parameter $w$ which may be time dependent.
Together, DM and DE form a ``dark sector" which makes up about 95\% of the present cosmic energy budget and which therefore dominates the cosmological dynamics. In the simplest case, following the $\Lambda$CDM paradigm, DM and DE are considered to be independent entities, governed by separate conservation laws.
The more general case is, however, not to exclude the possibility of a non-gravitational coupling between these both components which results in a richer structure of the dark sector.
Moreover, it has been demonstrated, that ignoring a potentially existing interaction may lead to an incorrect interpretation of cosmological
observations \citep{b1}. Since neither the physical nature of DE nor that of DM are known, these models are necessarily phenomenological.
Since they differ from the standard $\Lambda$CDM model they are useful to test potential deviations from the latter.
While for the homogeneous and isotropic background dynamics a lot of models do fit the observations, their different perturbation dynamics may serve to limit the number of seriously competing approaches.
There exists a large body of literature in which non-gravitational interactions between DE and DM are considered.
A subclass of these activities is devoted to models of DE which keep an EoS parameter $w=-1$ as in the standard model, but generalize the latter insofar as $\rho_{X}$ is admitted to be time dependent. These models are also called decaying $\Lambda$
models \citep{b2,b3,b4,b5,b6,b7,b8,b9,b10,b11,b12,b13,b14,b15,b16,b17,b18,b19,b20,b21,b22,b23,b24,b25,b26,b27,b28,b29,b30,b31,b32}.  Our aim here is to study in this context the perturbation dynamics of a model in which DM and DE interact in such a way that the ratio of their energy densities obeys a power-law in the scale factor, i.e. $\rho_{M}/\rho_{X} \propto a^{-\xi}$, where $\xi$ is a constant parameter.
Such model was proposed by Dalal et al. \citep{b33} to address the coincidence problem. Independently of whether or not one
considers this problem to be really a problem, the ansatz by Dalal et al. gives rise to a testable alternative cosmological
dynamics which contains the standard model as a limiting case. Various aspects of the background dynamics of this model have
been studied so far and were confronted with observational data \citep{b35,b38,b36,b37}.
This model is different from most other interacting models insofar as the interaction term is proportional to the product of the energy densities of DE and DM, i.e., the interaction is nonlinear.
Most approaches in the literature deal with an interaction linear in the energy density of one of the components  (see, e.g., \cite{gavela,pettorino,salvatelli}.
But in its original form its validity is restricted to the homogeneous and isotropic background.
The dynamics depends directly on the scale factor which is not a covariant quantity.
Here we present a covariant generalization of this model and complement previous investigations by a gauge-invariant perturbation analysis.
To this purpose the scale factor is replaced by a general, covariantly defined  length scale which under the conditions of homogeneity and isotropy reduces to
the scale factor. This allows us to establish a covariant and gauge-invariant perturbation theory on the basis of which we
calculate the matter power spectrum and discuss its dependence on the parameter $\xi$.

The paper is organized as follows. In Sec.~\ref{interacting} we present the basic framework for an interacting
system of two perfect fluids. Our scaling model is established in Sec.~\ref{model} where we also study its dynamics
in the spatially homogeneous and isotropic background. The perturbation analysis is the subject of Sec.~\ref{perturbations}. It provides us with an expression for the matter-density perturbation which is analyzed
and observationally tested in Sec.~\ref{observations}. A summary of the paper is given in Sec.~\ref{summary}.

\section{Interacting two-component system}
\label{interacting}

We assume the cosmic substratum to be dynamically dominated by a dark sector,
made of DM and DE.
The substratum as a whole is characterized by a conserved total perfect-fluid type energy-momentum
tensor
$T_{ik} = \rho u_{i}u_{k} + p h_{ik}$
with $T_{\ ;k}^{ik} = 0$.
Here, $h _{ik}=g_{ik} + u_{i}u_{k}$ and $g_{ik}u^{i}u^{k} = -1$. The quantity $u^{i}$ is the total four-velocity of the cosmic substratum and latin indices run from $0$ to $3$.
Splitting the conservation laws into their timelike and spacelike parts, we have
\begin{equation}
\rho_{,a}u^{a} +  \Theta \left(\rho + p\right) = 0\quad \mathrm{and} \quad \left(\rho + p\right)\dot{u}^{a} + p_{,i}h_{}^{ai} = 0\,,
\label{embtot}
\end{equation}
respectively, where $\Theta \equiv u^{a}_{;a}$.
Now we assume a split of $T_{ik}$ into a matter component (subindex M) and a dark energy component (subindex X),
according to $T^{ik} = T_{M}^{ik} + T_{X}^{ik}$
with ($A= M, X$)
\begin{equation}\label{TA}
T_{A}^{ik} = \rho_{A} u_A^{i} u^{k}_{A} + p_{A} h_{A}^{ik} \
,\qquad\ h_{A}^{ik} = g^{ik} + u_A^{i} u^{k}_{A} \,.
\end{equation}
An interaction between the components is covariantly characterized by
\begin{equation}\label{Q}
T_{M\ ;k}^{ik} = Q^{i},\qquad T_{X\ ;k}^{ik} = - Q^{i}\,.
\end{equation}
Then, the separate energy-balance equations are
\begin{equation}
-u_{Mi}T^{ik}_{M\ ;k} = \rho_{M,a}u_{M}^{a} +  \Theta_{M} \left(\rho_{M} + p_{M}\right) = -u_{Ma}Q^{a}\
\label{eb1}
\end{equation}
and
\begin{equation}
-u_{Xi}T^{ik}_{X\ ;k} = \rho_{X,a}u_{X}^{a} +  \Theta_{X} \left(\rho_{X} + p_{X}\right) = u_{Xa}Q^{a}\,.
\label{eb2}
\end{equation}
The four-velocities, in general different for both components, obey $g_{ik}u_{A}^{i}u_{A}^{k} = -1$. The quantities $\Theta_{A}$ are $\Theta_{A} \equiv u^{a}_{A;a}$. In the homogeneous and isotropic background we assume all four-velocities to coincide: $u_{M}^{a} = u_{X}^{a} = u^{a}$. The coupled momentum balances are
\begin{equation}
h_{Mi}^{a}T^{ik}_{M\ ;k} = \left(\rho_{M} + p_{M}\right)\dot{u}_{M}^{a} + p_{M,i}h_{M}^{ai} = h_{M i}^{a} Q^{i}\
\label{mb1}
\end{equation}
and
\begin{equation}
h_{Xi}^{a}T^{ik}_{X\ ;k} = \left(\rho_{X} + p_{X}\right)\dot{u}_{X}^{a} + p_{X,i}h_{X}^{ai} = - h_{X i}^{a} Q^{i}\,,
\label{mb2}
\end{equation}
where $\dot{u}_{A}^{a} \equiv u_{A ;b}^{a}u_{A}^{b}$.

It is convenient to decompose the source term $Q^{i}$ into parts proportional and perpendicular to the total four-velocity,
\begin{equation}
Q^{i} = u^{i}Q + \bar{Q}^{i}\,,
\label{Qdec}
\end{equation}
where $Q \equiv - u_{i}Q^{i}$ and $\bar{Q}^{i} \equiv h^{i}_{a}Q^{a}$ with $u_{i}\bar{Q}^{i} = 0$.

\section{The model and its background dynamics}
\label{model}
\subsection{The model}
Following \citet{b34}, we introduce a covariant length scale $l$ by
\begin{equation}\label{l}
\frac{\dot{l}}{l} \equiv \frac{1}{3}\Theta\ , \qquad \dot{l} \equiv l_{,a}u^{a}\,.
\end{equation}
Our aim is to consider the dynamics of a class of models for which the ratio of the energy densities of both components,
$r \equiv \frac{\rho_{M}}{\rho_{X}}$, behaves as a power of the length scale $l$,
\begin{equation}\label{r}
r = \frac{\rho_{M}}{\rho_{X}} \quad \Rightarrow\quad r = r_{0}l^{-\xi}\ ,
\end{equation}
where $\xi$ is a constant parameter and $r_{0}$ is the present value of the ratio $r$.
The  evolution of the ratio $r$ is then given
\begin{equation}\label{drTh}
\frac{\dot{r}}{r} \equiv \frac{r_{,a}u^{a}}{r} = - \frac{\xi}{3}\Theta\ .
\end{equation}
Only tensorial quantities were used here.
Relations (\ref{l}), (\ref{r}) and (\ref{drTh}) generalize a previous model introduced
by \citet{b33} which subsequently was studied in detail in \citet{b35,b36,b37,b38}. In its original form, this model
was restricted to the homogeneous and isotropic background dynamics.
Our covariant generalization relying on the use of the length scale (\ref{l}) opens the possibility to consider an
inhomogeneous perturbation dynamics as well.
In the background one has $\Theta = 3 H = 3 \frac{\dot{a}}{a}$, were $a$ is the scale factor of the Robertson-Walker
metric and  (\ref{r}) reduces to $r = r_{0}a^{-\xi}$ which defines the class of models considered in \citet{b33}.

\subsection{Background dynamics}

Assuming a pressureless matter component, in the homogeneous and isotropic background the balance equations (\ref{eb1}) and (\ref{eb2}) reduce to
\begin{equation}
\dot{\rho}_{M} + 3H \rho_{M} = Q \quad \mathrm{and }\quad
\dot{\rho}_{X} + 3H (1+w)\rho_{X} = - Q \,, \label{dotrhomx}
\end{equation}
respectively,
where $w\equiv \frac{p_{X}}{\rho_{X}}$ is the EoS parameter of the DE and
$Q$ is the background value of the general source term $Q$.
Combining the background ansatz $r = r_{0}a^{-\xi} \Rightarrow \dot{r} = - \xi H r$ with the balances (\ref{dotrhomx}) yields
\begin{equation}
 Q = -3H\left(\frac{\xi}{3} + w\right)\frac{\rho _{M}\rho _{X}}{\rho} = - 3H\rho _{M}\frac{\frac{\xi}{3} + w}
{r_0 a^{-\xi} + 1}.
\label{Q1}
\end{equation}
This relation determines the interaction that is necessary to generate a dynamics with $r = r_{0}a^{-\xi}$.
The interaction vanishes for the special cases $\xi = - 3w$. The $\Lambda$CDM model is recovered for
$\xi = 3$ and $w = -1$. Every combination $\frac{\xi}{3} + w \neq 0$ gives rise to an alternative, testable model.

It is worth noting that we are not starting with a phenomenological ansatz for a potential interaction and then study the resulting dynamics. Rather we assume a specific dynamics, the power-law behavior $r = r_{0}a^{-\xi}$ with the $\Lambda$CDM dynamics as well-defined special case, and then find the coupling that is necessary to produce such behavior.

We emphasize that the interaction (\ref{Q1}) is a nonlinear interaction. It depends on the product of the densities
of the interacting components. This makes it different from and obviously more ``realistic" than most of the other interactions between DM and DE considered in the literature, which are frequently just linear in the DE density.

In fact, systems with nonlinear interaction are not, in general, analytically solvable. The case under consideration here is a rare exception.
Notice that it is only in the limit $\rho _{M}\gg \rho _{X}$, equivalent to $\rho _{M}\approx \rho$, i.e. at high redshift that the interaction becomes approximately linear in $\rho _{X}$ and approaches the type of coupling dealt with, e.g., in \cite{gavela} and \cite{salvatelli}.

For a constant EoS parameter $w$ the matter-energy balance in (\ref{dotrhomx}) can be integrated,
\begin{equation}
\rho _{M} =
\rho _{M0}
\left[1+z \right]^{3 \left(1 + w \right)+\xi }
\left[\frac{1 + r_{0} \left(1+z \right) ^{\xi }}{1+r }\right]^{-1-\frac{3 w}{\xi }}\,,
\label{}
\end{equation}
where $z = \frac{1}{a} - 1$ is the redshift parameter.
The total energy density becomes
\begin{equation}
\rho  = \rho _{0}a^{-3}\left(\frac{r_{0} + a^{\xi}}{r_{0} + 1}\right)^{-\frac{3 w}{\xi }}\,,\qquad
\rho_{0} = \frac{r_{0}+1}{r_{0}}\rho _{M0}\,,
\label{}
\end{equation}
where $\rho_{0}$ is the present value of the energy density $\rho$.
Restricting ourselves to a universe with spatially flat sections, we obtain
the Hubble rate
\begin{equation}
\label{Ha}
H = H_{0}a^{-\frac{3}{2}}\left(\frac{r_{0} + a^{\xi}}{r_{0} + 1}\right)^{-\frac{3 w}{2\xi }}
\
\end{equation}
and the deceleration parameter $q = - 1 - \frac{\dot{H}}{H^{2}}$,
\begin{equation}\label{}
q = \frac{1}{2} +  \frac{3}{2}\frac{w}{r_{0}a^{-\xi} + 1}\ .
\end{equation}
The present matter fraction $\Omega_{M0}$ of the Universe is related to the ratio $r_{0}$ by
$\Omega_{M0} = r_{0}/(1+r_{0})$.
For the special case $p_{X} = - \rho_{X}$ we have (in the background)
\begin{equation}
T_{X}^{ik} = - \rho_{X}g^{ik}\qquad \mathrm{and }\qquad \dot{\rho}_{X}  = - Q\ ,
\label{Tx}
\end{equation}
equivalent to a \textbf{time-varying}  cosmological term.

\section{Perturbation dynamics}
\label{perturbations}

\subsection{General setup}

Let us denote first-order perturbations by a hat symbol.  Assuming the equality of all four velocities in the background, $u_{M}^{a} = u_{X}^{a} = u^{a}$, it follows from $g_{ik}u_{A}^{i}u_{A}^{k} = -1$ that the perturbed time components of the four-velocities are equal as well, i.e., $\hat{u}_{0} = \hat{u}^{0} = \hat{u}_{M}^{0} =\hat{u}_{X}^{0}  = \frac{1}{2}\hat{g}_{00}$.
Because of this property,
at first order, the expressions $\rho_{M,a}u_{M}^{a}$ and $\rho_{X,a}u_{X}^{a}$ in the energy balances (\ref{eb1}) and (\ref{eb2}), respectively, are
\begin{equation}\label{}
\left(\rho_{M,a}u_{M}^{a}\right)^{\hat{}} = \hat{\rho}_{M,a}u^{a}_{M} + \rho_{M,a}\hat{u}^{a}_{M}
= \dot{\hat{\rho}}_{M} + \dot{\rho}_{M}\hat{u}^{0}
\end{equation}
and
\begin{equation}\label{}
\left(\rho_{X,a}u_{X}^{a}\right)^{\hat{}} = \hat{\rho}_{X,a}u^{a}_{X} + \rho_{X,a}\hat{u}^{a}_{X}
= \dot{\hat{\rho}}_{X} + \dot{\rho}_{X}\hat{u}^{0}\ ,
\end{equation}
respectively. This implies that at first order
\begin{equation}\label{dderequal}
\rho_{M,a}u_{M}^{a} = \rho_{M,a}u^{a} \quad \mathrm{and} \quad \rho_{X,a}u_{X}^{a} = \rho_{X,a}u^{a}\
\end{equation}
are valid.
The timelike projections of the derivatives along the four velocities of the components coincide with the corresponding projections along the total four velocity.
In other words, there is only one timelike derivative. Obviously, this is no longer valid at higher orders.

Combining the balances (\ref{eb1}) and (\ref{eb2}) with (\ref{dderequal})
it follows that, up to first order,
\begin{equation}\label{xi=}
\frac{\xi}{3}\Theta = \Theta_{M} + \frac{u_{Ma}Q^{a}}{\rho_{M}} + \left[-\Theta_{X}\left(1+w\right)
+ \frac{u_{Xa}Q^{a}}{\rho_{X}}\right]\,.
\end{equation}
In the background ($\Theta_{M} = \Theta_{X} = \Theta$ and $u_{Ma}Q^{a} = u_{Xa}Q^{a} = u_{a}Q^{a} = -Q$) we recover relation (\ref{Q1}).
Equation (\ref{xi=}) will be crucial to determine the perturbed source in the following subsection.

In a next step we define the scalar
velocity potentials $v$, $v_M$ and $v_{X}$ for the spatial velocity perturbations by (Greek indices run from $1$ to $3$)
\begin{equation}\label{defpot}
\hat{u}_{\alpha} = v_{,\alpha}\,,\quad \hat{u}_{M\alpha} = v_{M,\alpha}\,,\quad \hat{u}_{X\alpha} = v_{X,\alpha}\,.
\end{equation}
Directly from the definition of $\Theta$ it follows that
\begin{equation}
\hat{\Theta} = \frac{1}{a^2}\left (\Delta v +\Delta \chi\right) -
3\dot{\psi} - 3 H\phi\,, \label{Thetaexp}
\end{equation}
where $\Delta$ is the three-dimensional Laplacian and where we have introduced the line element for scalar perturbations,
\begin{eqnarray}
\mbox{d}s^{2} = - \left(1 + 2 \phi\right)\mbox{d}t^2 + 2 a^2
F_{,\alpha }\mbox{d}t\mbox{d}x^{\alpha} +
a^2\left[\left(1-2\psi\right)\delta _{\alpha \beta} \right. \nonumber \\
+ \left.2E_{,\alpha
\beta} \right] \mbox{d}x^\alpha\mbox{d}x^\beta , \label{ds}
\end{eqnarray}
together with the abbreviation
\begin{equation}
\chi \equiv a^2\left(\dot{E} -F\right) \,.
\label{}
\end{equation}
Via the variable $F$ the spatial velocity components $\hat{u}^\mu$ are related to the potential $v$,
\begin{equation}
a^2\hat{u}^\mu + a^2F_{,\mu} = \hat{u}_\mu \equiv v_{,\mu} \,.
\label{}
\end{equation}
Analogous relations are valid for the velocity variables of the components.
Similarly to (\ref{Thetaexp}) one has ($A = X,M$)
\begin{equation}
\hat{\Theta}_{A} = \frac{1}{a^2}\left(\Delta v_{A} +\Delta \chi\right) -
3\dot{\psi} - 3 H\phi \,.
\label{Thetaexp1}
\end{equation}
For the differences $\hat{\Theta}_{A} - \hat{\Theta}$
it follows that
\begin{equation}\label{}
\hat{\Theta}_{A} - \hat{\Theta} = \frac{1}{a^{2}}\left(\Delta v_{A} - \Delta v\right)\,.
\end{equation}

According to the perfect-fluid structure of both the total energy-momentum tensor and the energy-momentum tensors of the components in (\ref{TA}), and with $u_{M}^{a} = u_{X}^{a} = u^{a}$ in the background, we have first-order energy-density perturbations
$\hat{\rho} = \hat{\rho}_{M} + \hat{\rho}_{X}$, pressure perturbations $\hat{p} = \hat{p}_{M} + \hat{p}_{X} = \hat{p}_{X}$
and
\begin{eqnarray}
\hat{T}^{0}_{\alpha} &=& \hat{T}^{0}_{M\alpha} + \hat{T}^{0}_{X\alpha}\nonumber\\
&&\Rightarrow
 \left(\rho + p\right)\hat{u}_{\alpha} = \rho_{M}\hat{u}_{M\alpha} + \left(\rho_{X} + p_{X}\right)\hat{u}_{X\alpha}.
\label{T0al}
\end{eqnarray}

\noindent
With the definitions in (\ref{defpot}) relation (\ref{T0al}) implies
\begin{eqnarray}\label{}
v_{M} - v&= &\left(1+w\right)\frac{\rho_{X}}{\rho + p}\left(v_{M} - v_{X}\right)\ , \nonumber \\
v_{X} - v& =& - \frac{\rho_{M}}{\rho + p}\left(v_{M} - v_{X}\right)\ .
\end{eqnarray}

\subsection{The perturbed source term}

From now on we shall restrict ourselves to the
case $p_{X} = - \rho_{X}$, equivalent to an EoS parameter $w=-1$, i.e., to perturbed vacuum energy.
The departure of the background dynamics from the $\Lambda$CDM model is then quantified by the difference
of the parameter $\xi$ from $\xi = 3$.
Under this condition
it follows from  (\ref{T0al}) that
\begin{eqnarray}
p_{X} = - \rho_{X} \quad\Rightarrow\quad \rho + p = \rho_{M} \quad \Rightarrow \quad\hat{u}_{M\alpha} = \hat{u}_{\alpha}\quad
\nonumber \\
\Rightarrow\quad v_{M} = v.
\label{ual}
\end{eqnarray}
Since the component $M$ is supposed to describe matter, it is clear from (\ref{T0al}) that the perturbed matter velocity $\hat{u}_{M\alpha}$ coincides with the total velocity perturbation $\hat{u}_{\alpha}$.
With $u^{n}_{M} = u^{n}$ up to first order, the energy balance in (\ref{eb1}) (correct up to first order) can be written as
\begin{equation}
\rho_{M,a}u^{a} = -  \Theta \rho_{M}  -u_{a}Q^{a}\,,\qquad (w=-1)\,.
\label{eb1u}
\end{equation}
On the other hand, the total energy balance (cf.~(\ref{embtot})) is
$\rho_{,a}u^{a} = - \Theta \left(\rho + p\right)$.
For the difference we find
\begin{equation}
\dot{\rho} - \dot{\rho}_{M} \equiv \left(\rho - \rho_{M} \right)_{,a}u^{a}
= u_{a}Q^{a}
\ .
\label{diffeb}
\end{equation}
Since, at least up to linear order, $\rho - \rho_{M} = \rho_{X}$, Eq.~ (\ref{diffeb}) is equivalent (up to the first order) to
\begin{equation}
\dot{\rho}_{X} \equiv \rho_{X,a}u^{a} = u_{a}Q^{a}
\,.
\label{drx}
\end{equation}
In zeroth order we consistently recover (\ref{Tx}).
The first order of Eq.~(\ref{drx}) is
\begin{equation}
\dot{\hat{\rho}}_{X} + \dot{\rho}_{X}\hat{u}^{0} = \left(u_{a}Q^{a}\right)^{\hat{}}
\,.
\label{diffepert}
\end{equation}
Notice that Eq.~(\ref{diffepert}) results from a combination of the total energy conservation and the matter energy balance. It has to be consistent with the DE balance  (\ref{eb2}). At first order, the latter becomes
\begin{equation}
\dot{\hat{\rho}}_{X} + \dot{\rho}_{X}\hat{u}^{0} = \left(u_{Xa}Q^{a}\right)^{\hat{}}
\ .
\label{eb2pert}
\end{equation}
Consistency then requires that
\begin{equation}
\left(u_{Xa}Q^{a}\right)^{\hat{}} = \left(u_{a}Q^{a}\right)^{\hat{}}
\ ,
\label{u2u}
\end{equation}
i.e., the projections of $Q^{a}$ along $u_{Xa}$ and along $u_{a}$ coincide. Explicitly,
\begin{equation}
\left(u_{a}Q^{a}\right)^{\hat{}} = \left(u_{a}u^{a}Q\right)^{\hat{}} = - \hat{Q}
\ .
\label{uQpert}
\end{equation}
Under the conditions (\ref{u2u}) and (\ref{uQpert}) relation (\ref{xi=}) becomes (for $w = -1$)
\begin{equation}\label{}
\frac{\xi}{3}\hat{\Theta} = \hat{\Theta}_{M}
-\hat{Q}\frac{\rho}{\rho_{M}\rho_{X}} - Q\left(\frac{\rho}{\rho_{M}\rho_{X}}\right)^{\hat{}}\,.
\end{equation}
Solving for $\hat{Q}$  we find, after some transformation,
\begin{equation}\label{hatQ}
\hat{Q}
= Q\left[\frac{\hat{\Theta}}{\Theta} + \delta_{M}\left(1 - r\right) + r\delta\right],\quad
\delta \equiv \frac{\hat{\rho}}{\rho},\quad \delta_{M} \equiv \frac{\hat{\rho}_{M}}{\rho_{M}},
\end{equation}
where $Q$ is the background expression (\ref{Q1}).
Since the combination of interest will be $\hat{Q}-Q\delta_{M}$, it is useful to rewrite (\ref{hatQ}) as
\begin{equation}\label{wh}
\widehat{\left(\frac{Q}{\rho_{M}}\right)} = \frac{Q}{\rho_{M}}\left[\frac{\hat{\Theta}}{\Theta} -r \left(\delta_{M} - \delta\right)\right] \,.
\end{equation}
With the perturbed source term explicitly known, we may now, in the following subsection, establish the basic set of perturbation equations.

\subsection{Basic set of perturbation equations}

To establish the basic set of perturbation equations it is
convenient
to introduce the gauge-invariant quantities
\begin{eqnarray}\label{defdc}
\delta^{c} = \delta + \frac{\dot{\rho}}{\rho}v \,, \qquad
\delta_{M}^{c} = \delta_{M} + \frac{\dot{\rho}_{M}}{\rho_{M}}v \,,  \nonumber \\
\delta_{X}^{c} = \delta_{X} + \frac{\dot{\rho}_{X}}{\rho_{M}}v \,, \qquad
\hat{p}_{X}^{c} = \hat{p}_{X} + \dot{p}_{X}v
\end{eqnarray}
as well as
\begin{equation}\label{Qc}
\hat{\Theta}^{c} = \hat{\Theta} + \dot{\Theta}v\  \qquad \mathrm{and} \qquad \hat{Q}^{c} = \hat{Q} + \dot{Q}v.
\end{equation}
The superscript $c$ stands for comoving. All the symbols have their physical meaning on comoving hypersurfaces $v=0$.
In terms of these gauge-invariant variables the total  energy and momentum conservations can be combined to yield (cf. \citep{b47})
\begin{equation}\label{balcomb}
\dot{\delta}^{c} - \Theta\frac{p}{\rho}\delta^{c} + \hat{\Theta}^{c}\left(1+\frac{p}{\rho}\right) = 0\,.
\end{equation}
The energy-density perturbations are coupled to the perturbations of the expansion scalar which are determined by the Raychaudhuri equation.
At first-order this equation becomes (cf. \citep{b47})
\begin{equation}\label{dThetacfin}
\dot{\hat{\Theta}}^{c} + \frac{2}{3}\Theta\hat{\Theta}^{c} + 4\pi G\rho\delta^{c}
+ \frac{1}{a^2}\frac{\Delta \hat{p}^{c}_{X}}{\rho + p} = 0\,.
\end{equation}
Combining Eqs.~(\ref{balcomb}) and (\ref{dThetacfin}),
changing to $a$ as independent variable ($\delta^{\prime} \equiv \frac{d \delta^{c}}{d a}$) and transforming
into the $k-$ space, the resulting equation for the total density perturbations is
\begin{eqnarray}
\delta^{c\prime\prime} + \left[\frac{3}{2}-\frac{15}{2}\frac{p}{\rho}+ 3\frac{p^{\prime}}{\rho^{\prime}}\right]\frac{\delta^{c\prime}}{a}\quad\quad\qquad\qquad\qquad &&\nonumber\\
 - \left[\frac{3}{2} + 12\frac{p}{\rho} - \frac{9}{2}\frac{p^{2}}{\rho^{2}} - 9\frac{p^{\prime}}{\rho^{\prime}}
\right]\frac{\delta^{c}}{a^{2}}
+ \frac{k^{2}}{a^{2}H^{2}}\frac{\hat{p}^{c}_{X}}{\rho a^{2}}
&=& 0.
  \label{dddeltak}
\end{eqnarray}
Because of the scale-dependent pressure perturbation term this is not a closed equation for $\delta^{c}$.
To clarify the role of this term in (\ref{dddeltak}) we introduce the sound speed $c_{s}$ in the rest-frame $v=0$ by $\hat{p}^{c}_{X} = c_{s}^{2}\hat{\rho}_{X}^{c}$.
The sound speed is considered here as a free parameter. In an interacting two-component system the sound does not, in general, propagate with the adiabatic sound speed $c_{ad}$, given by $c_{ad}^{2} = \dot{p}/\dot{\rho}$.
Then, the first-order energy balance for the matter component takes the form
\begin{equation}\label{dmc}
\dot{\delta}_{M}^{c}  + \hat{\Theta}^{c} + c_{s}^{2} \frac{\dot{\rho}_{M}}{\rho_{M}}\frac{\rho_{X}}{\rho_{M}}\delta_{X}^{c} = \widehat{\left(\frac{Q}{\rho_{M}}\right)}^{c}
\,,
\end{equation}
where
\begin{equation}\label{hatQc}
\widehat{\left(\frac{Q}{\rho_{M}}\right)}^{c} = \frac{\hat{Q}^{c}}{\rho_{M}} - \frac{Q}{\rho_{M}}\delta_{M}^{c} = \frac{Q}{\rho_{M}}\left[\frac{\hat{\Theta}^{c}}{\Theta} -r \left(\delta_{M}^{c} - \delta^{c}\right)\right]\,.
\end{equation}
Since $\hat{\rho}_{X}^{c} = \hat{\rho}^{c} - \hat{\rho}_{M}^{c}$, we realize that the pressure perturbations
can be written as
\begin{equation}\label{hatpc}
\frac{\hat{p}_{X}^{c}}{\rho_{M}} = c_{s}^{2}S_{M}\,,\qquad S_{M} \equiv D^{c} - \delta_{M}^{c}\,,
\end{equation}
where we have introduced the fractional quantity
\begin{equation}\label{}
D^{c} \equiv \frac{\rho^{c}}{\rho + p} =  \frac{1+ r}{r}\delta^{c} \qquad \Leftrightarrow \qquad \delta^{c} = \frac{r}{1+r}D^{c} \,.
\end{equation}
It becomes obvious that
via (\ref{hatpc}) the dynamics of $\delta^{c}$ in (\ref{dddeltak})
is coupled to $S_{M} = D^{c} - \delta_{M}^{c}$. To describe the dynamics of $S_{M}$ we couple equations (\ref{balcomb}) and (\ref{dmc}):
\begin{equation}\label{}
\dot{S}_{M} + \frac{Q}{\rho_{M}}D^{c} - \frac{\dot{\rho}_{M}}{\rho_{M}}\frac{\hat{p}_{X}^{c}}{\rho_{M}} = - \widehat{\left(\frac{Q}{\rho_{M}}\right)}^{c}
\,,
\end{equation}
with the right-hand side of this equation given by (\ref{hatQc}).
Eliminating $\hat{\Theta}^{c}$ in the expression (\ref{hatQc}) with the help of (\ref{balcomb}) provides us with
\begin{eqnarray}\label{}
\dot{S}_{M} + \frac{Q}{\rho_{M}}D^{c}
+ \left(\Theta -\frac{Q}{\rho_{M}}\right)\frac{\hat{p}_{X}^{c}}{\rho_{M}}
=
\frac{Q}{\Theta\rho_{M}}
\left[\dot{D}^{c}+ \left(\frac{Q}{\rho_{M}}  \right. \right.\nonumber \\ \left. \left. + \Theta\frac{r}{1+r}\right)D^{c}
- \Theta r\,S_{M}\right]
\ .
\end{eqnarray}
After some transformations the final equation for $S_{M}$ is
\begin{eqnarray}\label{Sfin}
S^{c\prime}_{M} + \frac{3}{1+r}\left[\left(\frac{\xi}{3} + r\right)c_{s}^{2}- r\left(\frac{\xi}{3} -1\right)\right]\frac{S^{c}_{M}}{a}\qquad \qquad &&\nonumber\\
= \left(1 - \frac{\xi}{3}\right)\frac{\delta^{c\prime}}{r}.&&
\end{eqnarray}
Equation (\ref{Sfin}) is coupled to the equation (\ref{dddeltak}) for $\delta^{c}$, which together with
(\ref{hatpc}) becomes
\begin{eqnarray}
\delta^{c\prime\prime} + \left[\frac{3}{2}-\frac{15}{2}\frac{p}{\rho}+ 3\frac{p^{\prime}}{\rho^{\prime}}\right]\frac{\delta^{c\prime}}{a}\quad\qquad \qquad\qquad \qquad\qquad &&\nonumber\\
- \left[\frac{3}{2} + 12\frac{p}{\rho} - \frac{9}{2}\frac{p^{2}}{\rho^{2}} - 9\frac{p^{\prime}}{\rho^{\prime}}
\right]\frac{\delta^{c}}{a^{2}}\quad \qquad\qquad&&\nonumber\\
+ \frac{k^{2}}{a^{2}H^{2}}c_{s}^{2}\frac{r}{1+r}\frac{S^{c}_{M}}{a^{2}}
= 0.&&
  \label{dddeltakS}
\end{eqnarray}
The coupled set of equations (\ref{Sfin}) and (\ref{dddeltakS}) is the main result of this paper.
The explicit structure of the coefficients is
\begin{eqnarray}\label{pdp}
\frac{p}{\rho} = - \frac{1}{1+r_{0}a^{-\xi}}\,,\quad
\frac{p^{\prime}}{\rho^{\prime}} = \frac{\frac{\xi}{3} - 1}{1+r_{0}a^{-\xi}}\,, \nonumber \\
H = H_{0}a^{-\frac{3}{2}}\left(\frac{r_{0} + a^{\xi}}{r_{0} + 1}\right)^{\frac{3}{2\xi}}\,.
\end{eqnarray}
The total EoS parameter approaches zero at high redshifts. The adiabatic sound speed square is positive for $\xi > 3$ and it is negative for $\xi < 3$.
For $\xi = 3$ one consistently recovers the $\Lambda$CDM model with $\frac{p^{\prime}}{\rho^{\prime}} = Q = 0$.
Only if the pressure perturbations of the dark energy are negligible, the equation for $\delta^{c}$ decouples from that for $S^{c}_{M}$.
\begin{figure}
{
\includegraphics[width=7cm]{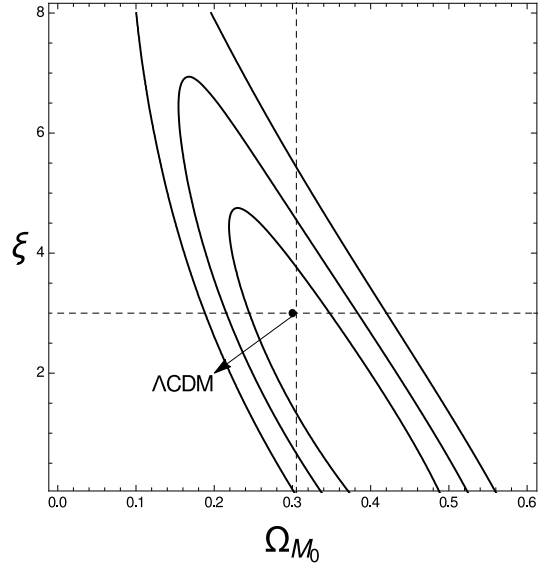}
}
\caption{Two-dimensional contour lines (1$\sigma$,
2$\sigma$ and 3$\sigma$ CL) in the $\Omega_{M0}$-$\xi$ plane, based on the JLA data set.
The point with the
best-fit values $\Omega_{M0}=0.30^{+0.04}_{-0.05}$ and $\xi=2.99^{+0.90}_{-1.45}$
for the scaling model is almost indistinguishable from the point that characterizes the $\Lambda$CDM model.}
\label{fig1}
\end{figure}
\subsection{Matter perturbations}
The set (\ref{Sfin}) and (\ref{dddeltakS}) for $S^{c}_{M}$ and $\delta^{c}$, respectively, describes the entire perturbation dynamics of the system. With this system solved, the matter perturbations $\delta^{c}_{M}$ are then obtained as the combination
\begin{equation}\label{dMc}
\delta^{c}_{M} = \frac{1+r}{r}\delta^{c} - S^{c}_{M}
 \,.
\end{equation}
To evaluate the set (\ref{Sfin}) and (\ref{dddeltakS}) we have to consider its behavior in the high-redshift limit.
Since $r\gg 1$ for $a\ll 1$ one has
\begin{equation}\label{}
\frac{p}{\rho} \ll 1\,,\qquad \frac{p^{\prime}}{\rho^{\prime}} \ll 1\,,\quad a^{2}H^{2}\gg H_{0}^{2}\qquad\quad (a\ll 1)\,.
\end{equation}
Under this condition Eq.~(\ref{dddeltakS}) reduces to
\begin{equation}\label{EdS}
\delta^{c\prime\prime} + \frac{3}{2}\frac{\delta^{c\prime}}{a}
- \frac{3}{2}\frac{\delta^{c}}{a^{2}}
= 0  \qquad\quad (a\ll 1)\,,
\end{equation}
which coincides with the corresponding equation for the Einstein-de Sitter universe.
It has the growing solution $\delta^{c} = c_{1}a$ where $c_{1}$ is a constant.
Equation (\ref{Sfin}) also decouples and reduces to
\begin{equation}\label{SMsmall}
S^{c\prime}_{M} + \left(3 - \xi + 3c_{s}^{2}\right)\frac{S^{c}_{M}}{a} = 0 \qquad\quad (a\ll 1)\,.
\end{equation}
It has the solution
\begin{equation}\label{Ssolearly}
S^{c}_{M} \propto a^{-\left(3-\xi + 3c_{s}^{2}\right)} \qquad\quad (a\ll 1)\,,
\end{equation}
which is constant exactly only in the $\Lambda$CDM limit $\xi = 3$ and $c_{s}^{2} = 0$.
It decays for $\xi < 3(1 + c_{s}^{2})$.
For the matter perturbations at $a\ll 1$ this implies
\begin{equation}\label{growthdel}
\delta^{c}_{M} \approx\delta^{c} \qquad\quad (a\ll 1)\,
\end{equation}
as expected for $r\gg 1$.

For $\xi > 3(1 + c_{s}^{2})$ the quantity $S^{c}_{M}$ may also grow. But as long as it remains close to
$\xi = 3$ the growth will be weaker than the growth of $\delta^{c}$ and (\ref{growthdel}) is still valid.

\begin{figure}
{
\includegraphics[width=7cm]{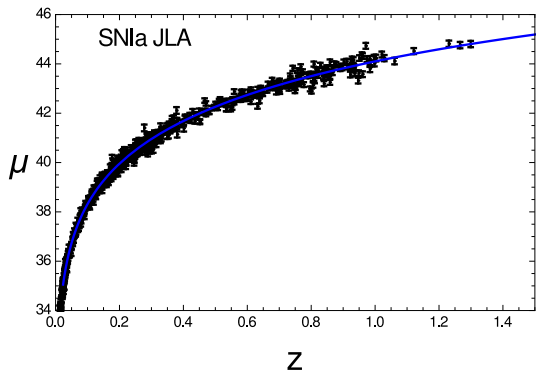}
}
\caption{Redshift dependence of the best-fit luminosity-distance modulus of our model compared with the data from
the JLA sample.}
\label{fig2}
\end{figure}

\section{Comparison with observational, results and discussions}
\label{observations}

Now we look for observational consequences of the model based on the ansatz (\ref{r}) both for the
homogeneous background dynamics and for structure formation.
Since any deviation from the standard model is accompanied by a non-gravitational coupling between DE and DM
this amounts to check the viability of the existence of such type of interaction in the dark
sector and to put limits on its strength.

As a first step we shall perform a  $\chi^2$-statistics both for the SNIa and for the LSS data in order to know, how the model is situated in the observational branch.
The $\chi^2$-statistics is based on the expression
\begin{equation}
\chi^{2}(\theta)=\transpose{\Delta y (\theta)}\mathbf{C}^{-1} \Delta y(\theta)
\label{chi2}
\end{equation}
with $\Delta y(\theta) = y_i-y(x_i;\theta)$  and the covariance matrix $\mathbf{C}$ of data $y_i$. The quantities $y_i$ represent the observational data
(in our case SNIa or LSS) which are compared with the theoretical predictions $y(x_{i}\vert\theta)$ with a set of parameters $\theta$.
Since we will consider the data as Gaussianly distributed the likelihood ${\cal{L}}$ is related to $\chi^2$ by
$\mathcal{L}\propto\exp\left(-\frac{\chi^{2}(\theta)}{2}\right)$.

Let us start with the 740 data points of the JLA sample \citep{b52}.
This updates a previous analysis in \citet{b36} based on the Constitution data set  \citep{b40}.
In this case $y$ represents the  luminosity-distance  modulus, which is theoretically
calculated as
\begin{equation}
\mu=5\log d_L(z)+\mu_0 \,,
\label{modulo}
\end{equation}
with $\mu_0=42,384-5\log h $, where
\begin{equation}
d_{L}=\left(z+1\right)H_0\int_{0}^{z}\frac{dz'}{H\left(z'\right)}\,,
\label{luminosidade}
\end{equation}
and $h$ is defined by $H_0 = 100  h \mathrm{km s^{-1} Mpc^{-1}}$. The Hubble rate $H(z)$ is given by eq. (\ref{Ha})
with $a$ replaced by $a = (1+z)^{-1}$. Observational data points of the luminosity-distance  modulus were calculated
using the relation \citep{b52}
\begin{equation} \label{muobs}
\mu_{obs} = m^*_B-(M_B-\alpha X_1+\beta C)\,,
\end{equation}
where $m^*_B$ corresponds to the observed peak magnitude in rest frame $B$ band and $\alpha$, $\beta$
and $M_B$ are nuisance parameters, $X_1$ is related to the time stretching of the light-curves,
and $C$ corrects the color at maximum brightness. In order to calculate completely $\mu_{obs}$
and its covariance matrix we followed the steps suggested in \cite{b52} and
used the JLA data \footnote{http://supernovae.in2p3.fr/sdss\_snls\_jla/ReadMe.html}.
\begin{figure}
{
\includegraphics[width=7cm]{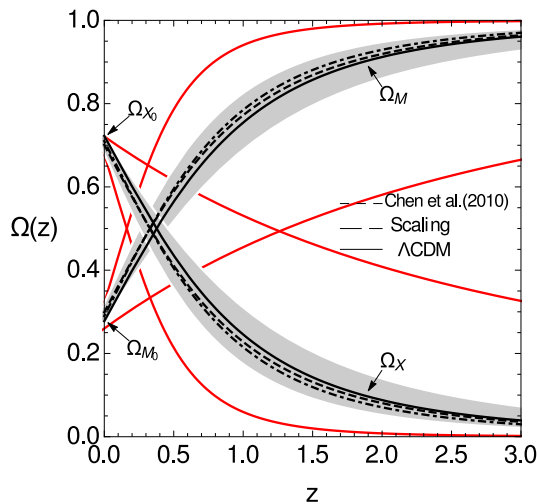}
}
\caption{Fractional densities of DE and DM as functions of the redshift.
The dot-dashed curves represent the best-fit values found in \citet{b36},
dashed curves correspond to the best-fit values of our model based on the JLA sample
and continuous curves denote the $\Lambda$CDM model. The red lines confine the 1$\sigma$ CL region of \citet{b36}, the gray area
represents that of our analysis.}
\label{fig3}
\end{figure}

In the most general case there are four free parameters:
$\theta = (h, \Omega_{M0},\xi,w)$. Then, marginalizing over $h$ and minimizing
the  $\chi^2$-function we could find the best-fit values for the three remaining free parameters.
However, the results based on the JLA data indicate an EoS parameter very close to $w=-1$.
Moreover, our perturbation analysis was performed for this case as well
 and thus hereinafter we will use  $w=-1$. Then we are left with the parameters $\xi$ and $\Omega_{M0}$.
On this basis we find the two-dimensional curves in the $\Omega_{M0}-\xi$ plane shown in Fig.~\ref{fig1}. The continuous curves represent the confidence
levels (CL) at  $1\sigma$, $2\sigma$ and $3\sigma$.
The point characterizes the best-fit values for our scaling model. It is almost indistinguishable from the point of best-fit
values for the $\Lambda$CDM model. Note that the values for $\xi$ are highly  degenerate. The curve in Fig. 2 shows the redshift dependence
of the best-fit luminosity-distance modulus
of our model compared with the data from the JLA sample.

\begin{figure}
{
\includegraphics[width=7cm]{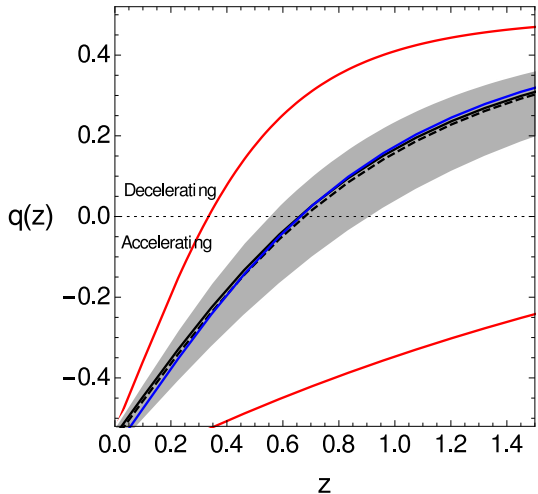}
}
\caption{Deceleration parameter as function of the redshift. The blue curve represents
the best-fit values in \citet{b36}, the dashed curve corresponds to the best-fit values of our model
for the JLA sample  and the continuous curve characterizes the $\Lambda$CDM model.
The red lines confine the 1$\sigma$ CL region of \citet{b36}, the gray area represents  that of our analysis.}
\label{fig4}
\end{figure}

In order to illustrate the consequences of the best-fit values of the model for the
background dynamics, we plot the fractional densities of DE
and DM and of the deceleration parameter as functions of the redshift
in Figures \ref{fig3} and \ref{fig4}, respectively. In these figures we also included the results of
a previous analysis in \citet{b36} (dot-dashed curves in  Fig.~\ref{fig3} and blue curve in Fig.~\ref{fig4}). The dashed curves represent
the best-fit values for our model, the solid curves those for the $\Lambda$CDM
model. The red lines confine the 1$\sigma$ CL region of \citet{b36}, the gray area represents  that of our analysis.
The error for the best-fit value of $\Omega_{M}$ is reduced by approximately 25\%, the error of the best-fit $\xi$ value by approximately  73\%
in relation to \cite{b36}.
Compared with the use of the Union 2.1  sample \citep{b39},
the error of $\Omega_{M}$ is reduced by about 5\%  and  the error of $\xi$ by about 39\%. Note that the behavior
of  the scaling model is practically the same as that of the $\Lambda$CDM model, i.e.,
the current SNIa samples  can not discriminate between these models. The same indistinguishability
happens  for the deceleration parameter as seen in Fig.~\ref{fig4} where we plotted $q(z)$
for our model, for the model in \citet{b36} and for the $\Lambda$CDM  model.
Again, the gray color indicates the region at 1$\sigma$ CL.

\begin{figure*} 
  \subfloat[]{ 
   \includegraphics[width=6.1cm,height=4.8cm]{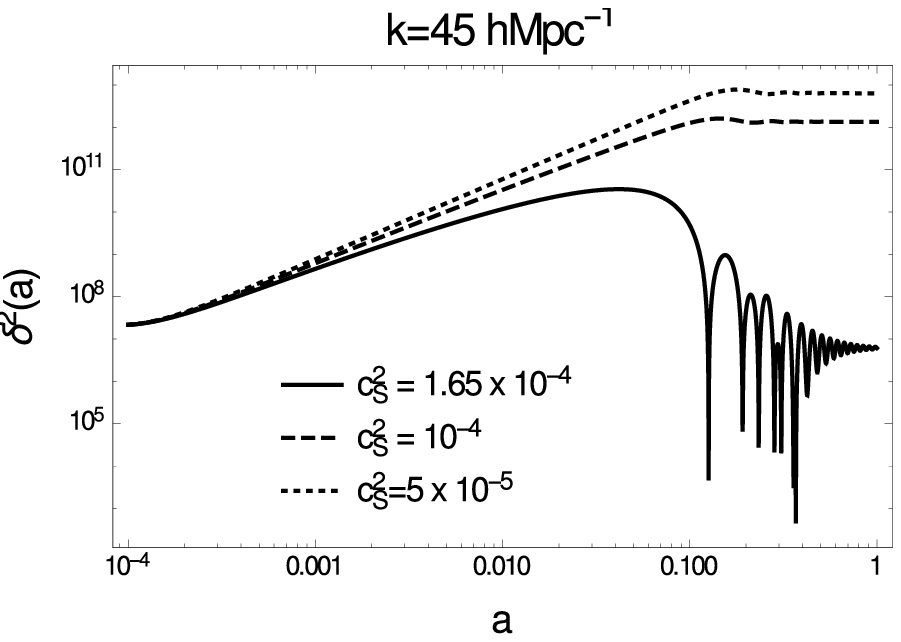}
  } 
  \subfloat[]{ 
    \includegraphics[width=6.1cm,height=4.8cm]{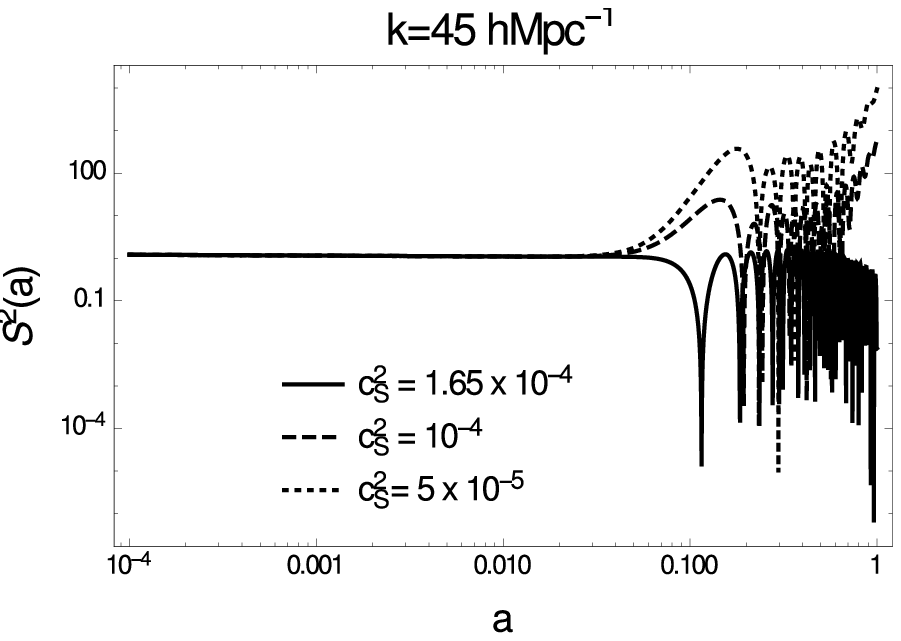}
  } 
  \subfloat[]{ 
    \includegraphics[width=6.1cm,height=4.8cm]{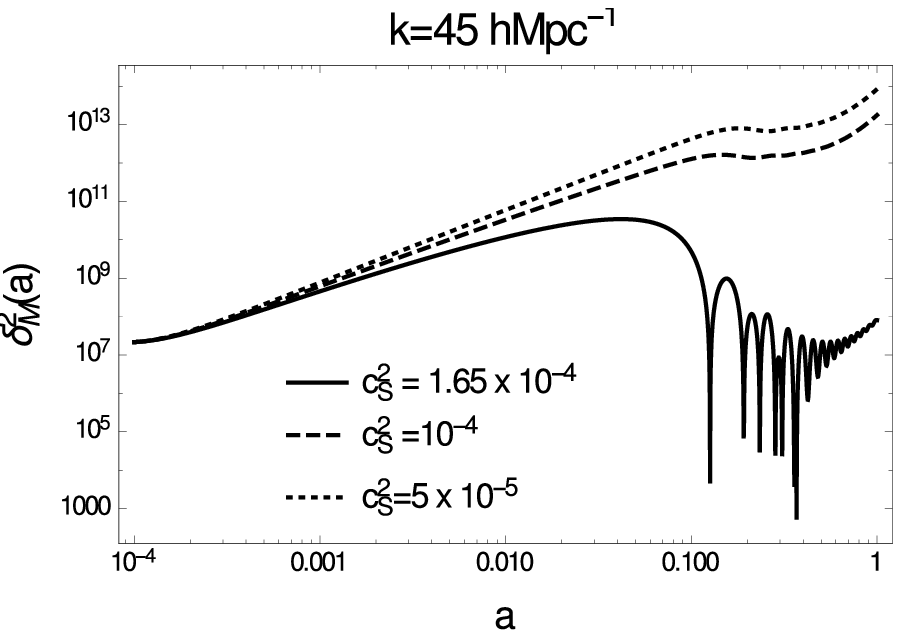}
  } 
  \\ 
  \subfloat[]{ 
    \includegraphics[width=6.1cm,height=4.8cm]{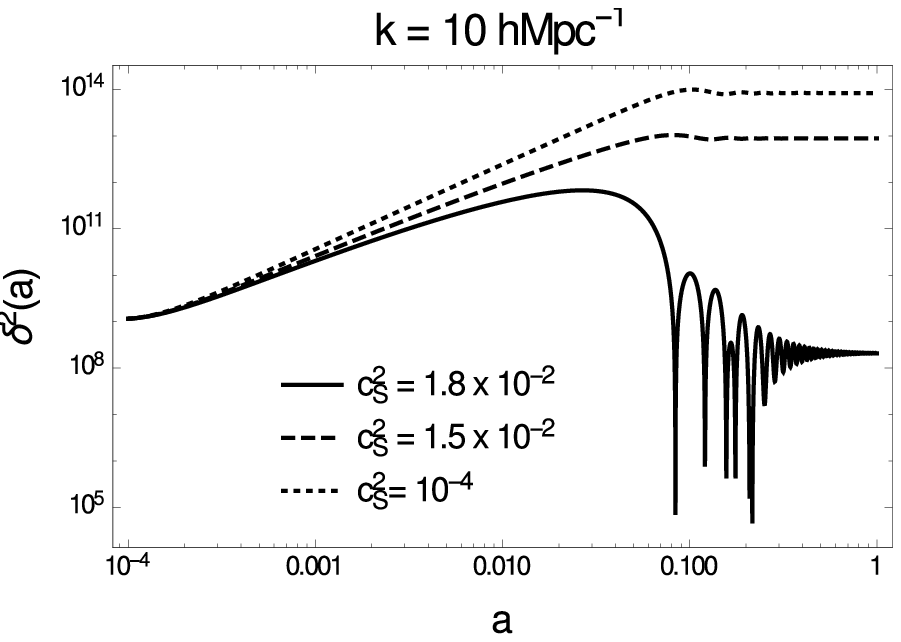}
  } 
  \subfloat[]{ 
    \includegraphics[width=6.1cm,height=4.8cm]{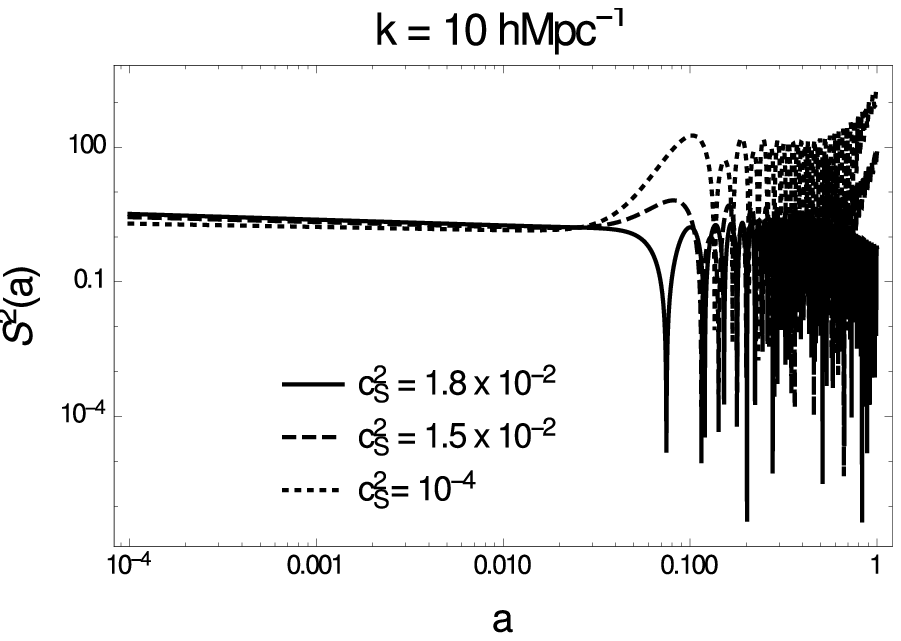}
  } 
  \subfloat[]{ 
    \includegraphics[width=6.1cm,height=4.8cm]{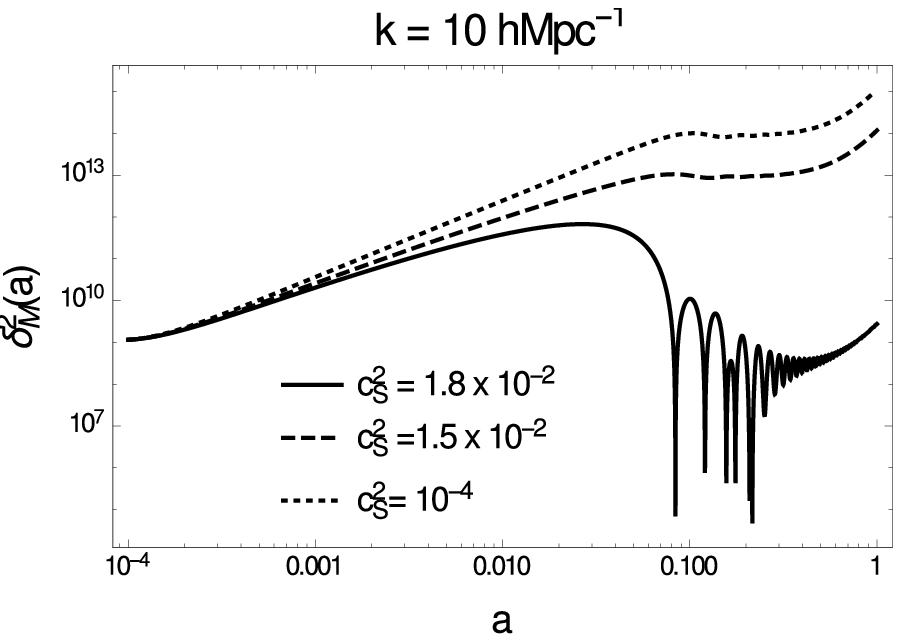}
  } 
  \\ 
  \subfloat[]{ 
    \includegraphics[width=6.1cm,height=4.8cm]{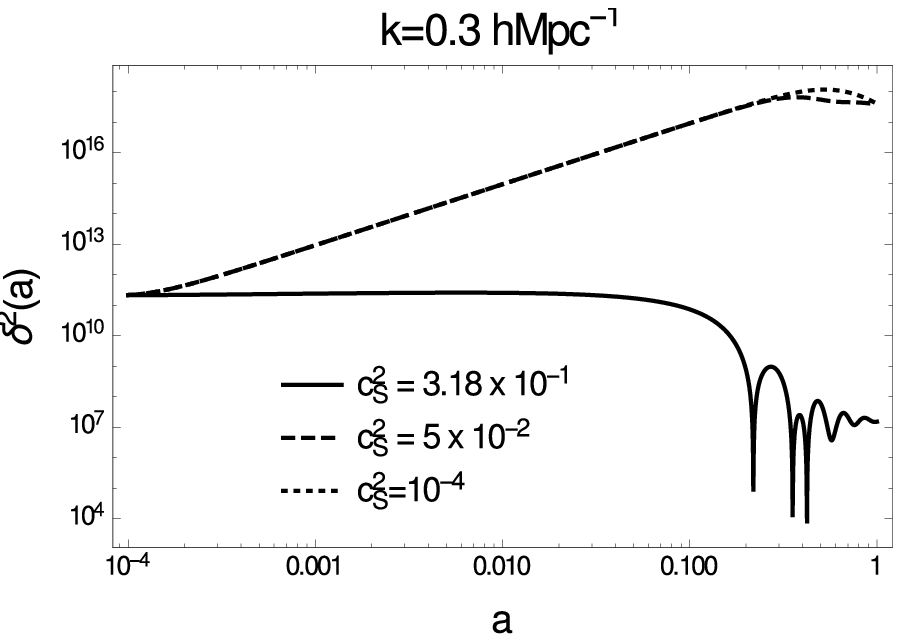}
  } 
  \subfloat[]{ 
    \includegraphics[width=6.1cm,height=4.8cm]{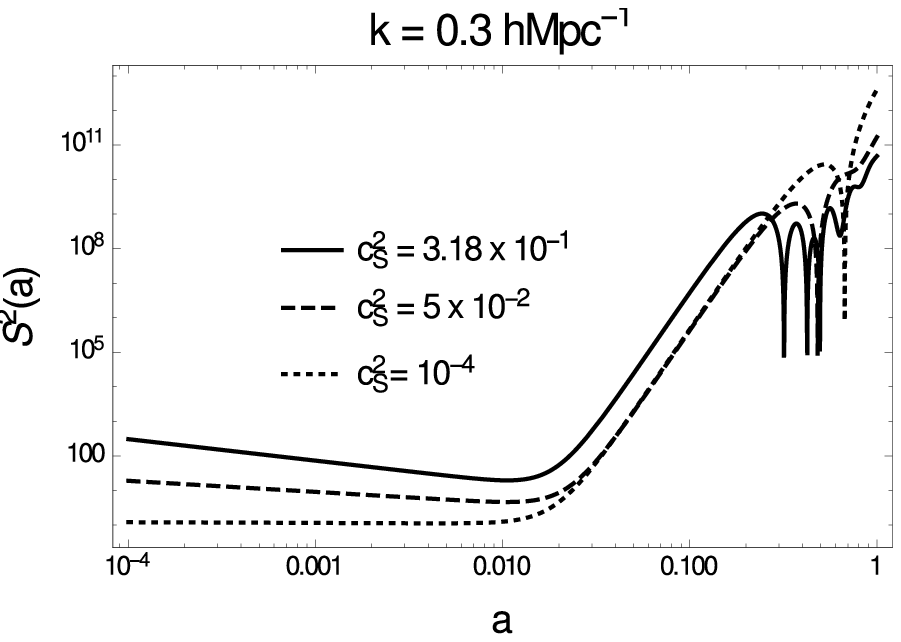}
  } 
  \subfloat[]{ 
    \includegraphics[width=6.1cm,height=4.8cm]{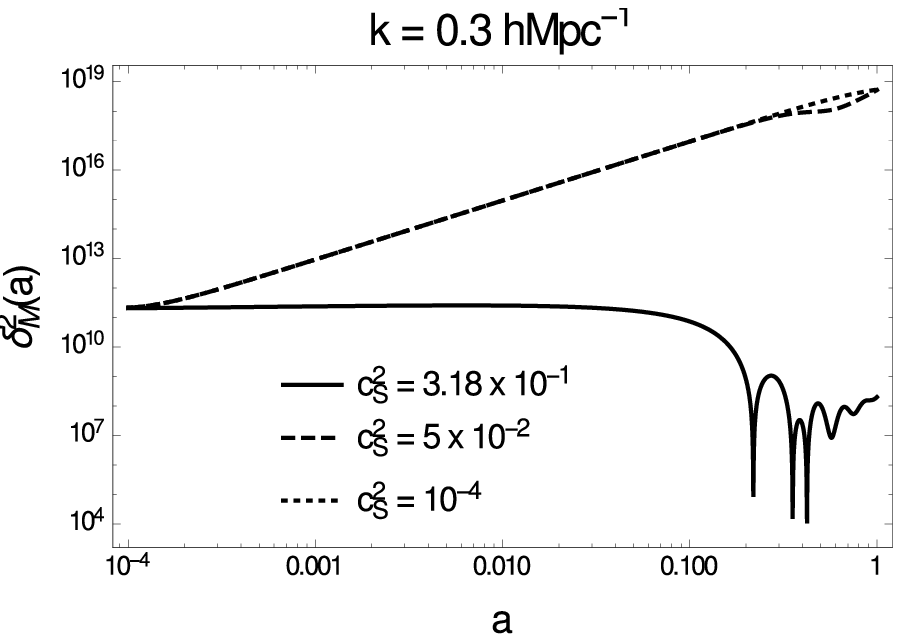}
  } 
 \\
  \caption{Total density contrast $\delta$, relative density contrast $S_{M}$ and matter density contrast  $\delta_M$ as functions of the scale factor for three different scales. Figures 5a, 5b and 5c represent the curves for $k=45 ~h\mathrm{Mpc}^{-1}$, figures 5d, 5e and 5f those for $k=10~h\mathrm{Mpc}^{-1}$ and figures 5g, 5h and 5i
those for $k=0.3\, h\mathrm{Mpc}^{-1}$. We have assumed here $\Omega_{M0}=
0.275$ and $\xi =2.99$.  In all cases high values of the sound speed lead to
oscillations while smaller values do not. (Note that we have omitted here the superscripts c.)}
\label{fig5}
  \end{figure*}

Now let us perform an analysis of the perturbation dynamics. Here we have the sound speed square $c^2_S$ as an additional parameter.
To get an idea of the role of this parameter
we  solved the system (\ref{Sfin}) and (\ref{dddeltakS}) for different values of $c^2_S$ with initial conditions provided by the limits
of the model for $a<<1$. In this limit eq.~(\ref{dddeltakS}) reduces to eq.~(\ref{EdS}), while the initial condition for eq.~(\ref{Sfin}) is found from eq.~(\ref{Ssolearly}) which governs the evolution of $S_M$
at early times. This procedure to choose initial conditions is similar to that described in more detail
in \citet{b45,b46}. The results for the integration of the system (\ref{Sfin})-(\ref{dddeltakS})
 for a typical case, we have chosen here $\Omega_{M0}=0.275$ and $\xi =2.99$,
are shown in Fig.~\ref{fig5} which visualizes the  evolution of total density contrast,
the relative  density contrast and the matter
density contrast  as functions of the scale factor for three different
scales.
Figures 5a, 5b and 5c present the curves for $k=45 \,hMpc^{-1}$,  figures 5d, 5e
and 5f those for $k=10~h\mathrm{Mpc}^{-1}$ and figures 5g, 5h and 5i those
for $k=0.3~h\mathrm{Mpc}^{-1}$. Similar plots can be obtained for different values of $\Omega_{M0}$ and $\xi$.

Several features of Fig.~\ref{fig5} are worth discussing. For example, according eq.~(\ref{Ssolearly}) the initial conditions  for $S_M$  depend on the sound velocity square $c^2_S$. However, the behavior
of $S_M$ is similar for all the cases in Fig.~\ref{fig5}.  It has constant values
(or almost constant values for $k=0.3~h\mathrm{Mpc}^{-1}$) in early times and starts  oscillating between $a\approx 0.08$ and $a\approx 0.12$.
The amplitude of these oscillations is small and  contributes only marginally to the matter density contrast $\delta_M$ (see eq.~(\ref{dMc})). Thus, $\delta$ and $\delta_M$ (we have omitted the superscripts c in this figure) show a very similar behavior, i.e., relation  (\ref{growthdel}) remains approximately valid. The evolution of both the total density contrast $\delta$ and of the
matter density contrast $\delta_M$ depends crucially on $c^2_S$. Our calculations show that for higher values
of the sound velocity square $c^2_S$ there appear oscillations both in $\delta$ and in $\delta_M$ at about $a\approx 0.1 - 0.2$. At values of $c^2_S$ lower than a threshold value $c^2_{S0}$, i.e., for $c^2_{S} < c^2_{S0}$ the oscillations disappear.
We confirmed numerically that $c^2_{S0}$ is inversely proportional to $k$. For example, for $k=45~h\mathrm{Mpc}^{-1}$ we have $c^2_{S0}= 1.7\times 10^{-4}$ (see figures 5a, 5b and 5c), for  $k=10~h\mathrm{Mpc}^{-1}$ the threshold changes to
$c^2_{S0} = 1.8\times 10^{-2}$ (see figures 5d, 5e and 5f) and for $k=0.3~h\mathrm{Mpc}^{-1}$ the threshold value is
$c^2_{S0} = 3.2 \times 10^{-1}$ (see figures 5g, 5h and 5i).

\begin{figure}
{
\includegraphics[width=7cm]{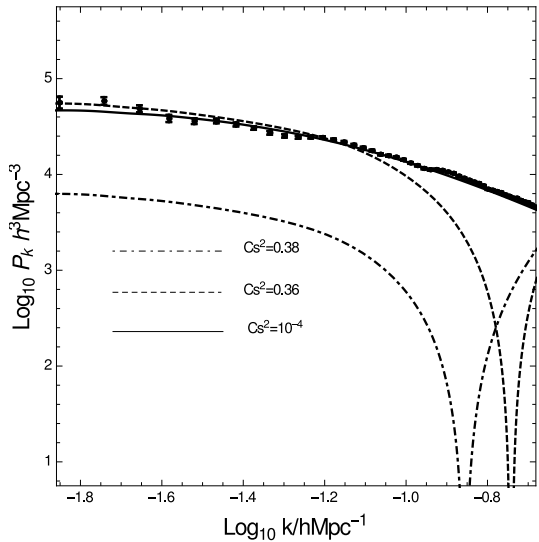}
}
\caption{Matter power spectrum for different values of the sound velocity.
There are oscillations for large values of $c^2_S$ but not for smaller values.}
\label{fig6}
\end{figure}

  Since with the solution of our basic system the matter density contrast $\delta_{M}(a)$ is known according to relation (\ref{dMc}), we can construct the matter power spectrum
($P_k$) and study the influence of $c^2_S$ on this spectrum.
In Fig.~\ref{fig6} we show the  matter power spectrum
for different values of the sound velocity. As to be expected, also the spectrum exhibits oscillations for larger values of
$c^2_S$. Such type of behavior is similar to what is known from unified models of the dark sector, notably from Chaplygin-gas
models. Since oscillations of the matter distribution are not observed, those models have temporarily fallen out of favor
\citep{b48}. However, non-adiabatic pressure perturbations may reduce the effective sound speed to very small values which may cure the problem of
oscillations \citep{b47,b49}. As a consequence, unified models continue to be discussed as potential alternatives to the $\Lambda$CDM model.
In our case $c^2_S$ is a free parameter and we shall discard models for which $c^2_S$ is larger than the mentioned threshold value $c^2_{S0}$.

Now, the $y_{i}$ in the expression (\ref{chi2}) represent the LSS DR9 data \citep{b50}, while $y$ is the theoretically calculated matter power spectrum $P_k$.
The set of free parameters is $\theta=(c^2_S,\Omega_{M0},\xi)$.
We have used the publicly available DR9 data covariance matrix and restricted  the wavenumber to the range $0.002 h \mathrm{Mpc}^{-1}<k< 0.2 h\mathrm{Mpc}^{-1}$ for which a linear approximation seems appropriate.

\begin{figure}
{
\includegraphics[width=7cm]{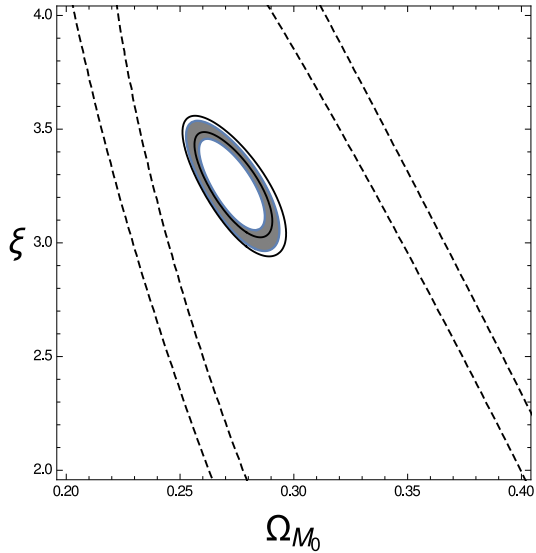}
}
\caption{$\Omega_{M0}$-$\xi$ plane for the scaling model.
Continuous contour lines represent the
$1\sigma$ and $2\sigma$ CL based on the DR9 data sample. For comparison we have also included the results for the JLA sample (dashed lines).
The grey region  represents the $1\sigma$ and $2\sigma$ CL for the
joint analysis JLA+DR9.}
\label{fig7}
\end{figure}

\begin{figure}
{
\includegraphics[width=7cm]{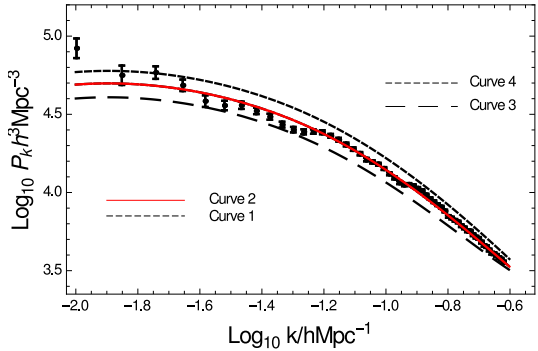}
}
\caption{Matter power spectrum with $\Omega_{M0} = 0.27$ for several values of $\xi$, including the
best-fit value, for the DR9 data.}
\label{fig8}
\end{figure}

 To avoid the mentioned unwanted and non-observed oscillations
very small values of $c^2_S$ are required. For our calculations we take $c^2_S\sim 10^{-4}$.
Such value will ensure  the absence of  oscillations until  $k\sim 45\, h\mathrm{Mpc}^{-1}$,
equivalent to a scale of $\sim 0.02\, h^{-1} \mathrm{Mpc}$. Hence, we continue
our analysis with only two free parameters, namely $\Omega_{M0}$ and $\xi$.
Minimizing the  $\chi^2$-function under this condition we find the best-fit values visualized in the $\Omega_{M0}$-$\xi$
plane of Fig.~\ref{fig7}. The continuous contour lines represent the
$1\sigma$ and $2\sigma$ confidence levels.
We performed  also a joint analysis by using the combined SNIa + LSS data where
$\chi^2=\chi^2_{SNIa}+\chi^2_{LSS}$. The results are presented in Fig.~\ref{fig7} as well, where
the grey color represents the region between the $1\sigma$ and $2\sigma$  CLs. The best-fit values
for the combined analysis are summarized in Table 1.
There is a tendency for $\xi > 3$, i.e., for an energy transfer from DM to DE.
While the JLA data alone show a big degeneracy, the latter is largely removed by the large-scale-structure data.
In Fig.~\ref{fig8} we present four different curves.
Curve 1 is the best-fit model for the DR9 data set. It has $\Omega_m \sim 0.27$ , $\xi\sim 3.26$ and $\chi^2_{\nu,DR9}=1.20$.
Curve 2 represents the best-fit model for the combined DR9+JLA data.
 The best-fit values are $\Omega_m \sim 0.27$ , $\xi\sim 3.25$ and  $\chi^2_{\nu,DR9+JLA}=1.16$.
 Note that curve 1 and curve 2 are indistinguishable.
To countercheck these results we have also chosen two parameter combinations
 from outside the  $2$CL solid  contour: curve 3 is based on  $\Omega_m \sim 0.30$  and $\xi\sim 3.8$ and has
a   $\chi^2_{\nu,DR9+JLA}=2.14$ for the joint DR9+JLA data.  The same model has a $\chi^2_{\nu,DR9}=2.68$ when only DR9 data
are considered. Curve 4 is constructed with  $\Omega_m \sim 0.23$ and  $\xi\sim 2.85$ and has a  $\chi^2_{\nu,DR9+JLA}=2.49$ when the combined DR9+JLA data sets are used. 
For DR9 data only one has
$\chi^2_{\nu,DR9}=3.09$. The fit to these values outside the $2\sigma$ contour is
indeed considerably worse which supports our analysis.

\begin{table*}
 \centering
 \begin{minipage}{140mm}
  \caption{Best-fit values for  the different tests of the scaling model with
  errors at 1$\sigma$ CL. The source term $Q$ is given in units of $H^3_0 / 8\pi G$.}
  \begin{tabular}{p{0.30\linewidth}p{0.15\linewidth}p{0.15\linewidth}p{0.10\linewidth}p{0.15\linewidth}}
    \hline
        Data & $\Omega_{M0}$ & $\xi$ & $\chi^2_\nu$ &  $Q$\\
    \hline
    JLA &  $0.30^{+0.04}_{-0.05}$ & $2.99^{+0.90}_{-1.45}$ & $1.18$ & $+0.01^{+0.97}_{-0.61}$ \\
    DR9 & $0.27^{+0.02}_{-0.01}$ & $3.26^{+0.16}_{-0.16}$ & $1.20$ &$-0.15^{+0.09}_{-0.11}$\\
    DR9+JLA & $0.27^{+0.01}_{-0.01}$ & $3.25^{+0.15}_{-0.15}$ & $1.16$ &$-0.15^{+0.09}_{-0.09}$ \\
        \hline
  \end{tabular}
  \end{minipage}
\end{table*}
Finally, with the known best-fit values we can infer the upper and lower
bounds for the interaction strength. These values are presented in the outer right column of Table 1.
As already mentioned, there is a  tendency to values $\xi > 3$ which corresponds to $Q<0$, equivalent to a transfer
of energy from DM to DE.  Qualitatively, this is in accordance with the results
in \cite{gavela} and \cite{salvatelli} which are based on an interaction linear in the DE energy density.
On the other hand, thermodynamic considerations prefer $Q$ to be positive
\citep{b51}.

\section{Conclusions}
\label{summary}

With the intention to quantify and, possibly, to soften the coincidence problem, \citet{b33}
introduced a phenomenological parameter $\xi$ that governs the dynamics of the ratio of the energy densities of DM and DE.
Independently of whether one takes the coincidence problem seriously, the resulting cosmological dynamics represents a simple, testable modification of the standard model.

Starting point of this approach is not, as usual, an expression for the interaction between DM and DE but the mentioned dynamics of the energy-density ratio which can be compared with that of the $\Lambda$CDM model in a transparent manner. Any deviation from the standard dynamics is then  traced back to a specific coupling between DM and DE.
Different from most models discussed so far, this coupling is nonlinear, i.e., the interaction quantity $Q$ contains a product of the energy densities of DM and DE. Most other interacting models are just linear in the DE density.
For $\xi = 3$ and $w=-1$ the $\Lambda$CDM model is recovered. Any combination
$w + \xi/3 \neq 0$ corresponds to a non-gravitational interaction between DM and DE. Any value $\xi <3$ is
considered to make the coincidence problem less severe.  This model was tested against observational data in
\citep{b36,b38,b33,b37}.
However, in its original form and in the studies so far its validity  is restricted to the homogeneous and isotropic background dynamics.
Here we generalized this model to enable a study of the perturbation dynamics as well.
To this purpose we replaced the scale factor in the model defining relation by a more general covariantly defined length scale which reduces to the scale factor in the appropriate limit.
We performed a gauge-invariant first-order perturbation analysis which enabled us to obtain the fractional matter density perturbations as a combination from a coupled system of equations for the total and relative perturbations.
Like in other DE models, the effective non-adiabatic sound velocity has to be smaller than a certain
 threshold value to avoid (unobserved) oscillations in the matter perturbations.
Within a $\chi^{2}$ analysis and focusing on the case of perturbed vacuum energy $w=-1$, the resulting matter power spectrum was confronted with data from the SDSS DR9 survey.
We studied the dependence of the power spectrum on the values of the parameter $\xi$.
The results for the background dynamics were improved considerably compared with a previous study.
Using the JLA data alone results in a large degeneracy in the $\Omega_{M0}$-$\xi$ plane which does not allow for a definite conclusion concerning the sign of the interaction.
This degeneracy is substantially reduced by the DR9 large-scale-structure data which prefer $\xi \approx 3.25$. This corresponds to an energy transfer from DM to DE. 
The $\Lambda$CDM model with $\xi = 3$ is compatible with our analysis at the $2\sigma$ confidence level.

Finally we remark that for a more realistic study of matter clustering on smaller scales in the presence of DE and for a better understanding of the role of $c^2_S$ in structure formation a nonlinear treatment is necessary.
Processes as virialization and spherical collapse which were outside the scope of the present paper have
necessarily to be considered. We hope to come back to this in future work.
Further constraints on our nonlinear model are expected from CMB data. This is currently under investigation.

\section*{Acknowledgments}
ARF acknowledges support from CAPES (Brazil).
WSHR is thankful to FAPES by the grant (BPC No 476/2013)  under which this work was carried out.
WZ was supported by CNPq and FAPES. We thank the anonymous referee for constructive suggestions.

\label{lastpage}

\end{document}